\documentclass[preprint,12pt]{elsarticle}

\usepackage{amssymb}
\usepackage{subfig}
\usepackage{color}%
\usepackage{soul}%

\usepackage{setspace}
\journal{Nuclear Instruments and Methods B}

\begin{document}

\begin{frontmatter}



\title{Dynamic screening and energy loss of antiprotons \\ colliding with excited Al clusters}


\author[rvt]{Natalia E. Koval\corref{cor1}}
\ead{natalia$\_$koval$@$ehu.es}
\cortext[cor1]{Corresponding author, tel: (+34) 943 01 87 59}
\author[rvt,focal]{Daniel S\'{a}nchez-Portal}
\ead{sqbsapod$@$ehu.es}
\author[els]{Andrey G. Borisov}
\ead{andrei.borissov$@$u-psud.fr}
\author[rvt,focal]{Ricardo D\'{i}ez Mui\~{n}o}
\ead{rdm$@$ehu.es}
\address[rvt]{Centro de F\'{\i}sica de Materiales
CFM/MPC (CSIC-UPV/EHU), \\ Paseo Manuel de Lardizabal 5,
20018 San Sebasti\'an, Spain \\}
\address[focal]{
Donostia International Physics Center DIPC, \\
Paseo Manuel de Lardizabal 4, 20018 San Sebasti\'an,
Spain \\}
\address[els]{
Institut des Sciences Mol\'eculaires d'Orsay, ISMO,
\\ Unit\'e de Recherches CNRS-Universit\'e Paris-Sud UMR 8214, \\
B{\^a}timent 351, Universit\'e Paris-Sud, F-91405 Orsay Cedex, France}

\begin{abstract}
We use time-dependent density functional theory to calculate the energy loss of an antiproton colliding with a small Al cluster
previously excited. The velocity of the antiproton is such that non-linear effects in the electronic response of the Al cluster are
relevant. We obtain that an antiproton penetrating an excited cluster transfers less energy to the cluster than an antiproton penetrating
a ground state cluster. We quantify this difference and analyze it in terms of the cluster excitation spectrum.

\end{abstract}

\begin{keyword}
Energy loss \sep TDDFT \sep Metal cluster

\end{keyword}

\end{frontmatter}
\newpage
% \linenumbers

\section{Introduction}
\label{Intro}
\

A charged particle moving across a metallic target is able to create electronic excitations in the medium at the expense of its own
kinetic energy. Research on this phenomenon has been broad in condensed matter physics and materials science because of its relevance in
various fundamental and applied topics, such as radiation damage, medical physics, and ion sputtering. 

A key point in the theoretical analysis of the slowing down of charged particles in metals is the intensity of the perturbation that the
moving particle introduces in the medium.
For a particle of charge $Q$ moving with a velocity $\upsilon$ in a standard metal, the perturbation strength can be roughly characterized
by the Sommerfeld parameter $\eta=Q/\upsilon$ \cite{nagy93}. If such ratio is small, $\eta << 1$, linear theory is naturally applied and
accurate results for the particle energy loss are found. If $\eta >> 1$ and $\upsilon$ is much smaller than the typical velocities of the
electrons in the medium, various non-perturbative methodologies have been successfully applied \cite{Echenique86}. In between these two
cases, in the regime of intermediate velocities, accurate descriptions of the energy loss process are much more involved because
quasistatic or perturbative approximations break down even for unit-charge projectiles. Only recently calculations based on time-dependent
density functional theory (TDDFT) \cite{Baer04,Quijada1,Quijada2,kras2007,pruneda2007,correa2012,zeb2012,castro2012} have shown its
potential to close this gap. 

The achievements of TDDFT in the non-linear description of electronic excitations pave the way to answer new and challenging questions in
the field. In the traditional description of energy loss processes, the target is always considered as initially in its ground state.
However, the energy lost by a travelling charge in a metallic medium should be affected by the electronic state in which the target is.
Otherwise said, if electronic excitations have been already created in the system, the electronic response to the incident perturbation,
and consequently the energy loss, will be different. In this work we try to quantify this difference for the particular case of a point
charge crossing metallic clusters of a few \AA \, size.

The experimental investigation of the excitation and ionization of neutral metal clusters by collision with
positively or negatively charged particles has been intensive. In particular, the ionization of metal clusters
by low energy singly and multiply charged ions and protons~\cite{Chandezon95,Daligault02} and the ionization of neutral metal clusters
by slow electrons~\cite{Kasperovich99,Kasperovich00-1,Kasperovich00-2,Halder12} have been studied. Description of such processes from
the theoretical point of view is incomplete and requires further investigation. In our work we study the collision of an Al
cluster with a slow negative point charge (an antiproton). The choice of the antiproton as a projectile allows us to avoid complications
related to the electron capture by the cluster if the projectile is an electron and the electron capture by the projectile if the latter is an ion.
Our goal is to identify the distinct effects that arise in the dynamic screening and the projectile energy loss when the metallic target has been
previously excited by a preceding projectile. In spite of the fact that our model is simplified, the results of our study can contribute
to the understanding of the
fundamentals of the dynamical processes during collision of charged particles with metallic clusters.

We perform an explicit time propagation of the electronic state of the system
using TDDFT and evaluate the energy lost by the charge when crossing ground-state clusters. We compare this quantity with the amount of
energy lost when the projectile crosses a cluster excited from a previous collision. We show that the difference is appreciable and
give the explanation of this change as a consequence of the excited state of the cluster as well as of the emission of
electronic charge from the excited cluster. 

Non-linear effects in the excitation of metal clusters have been previously analyzed with TDDFT. In particular, electron dynamics in
clusters under intense laser fields are an active hot topic of research \cite{Fennel2010} because of the possibilities offered to explore
and control ultrafast processes. The resonance energy of collective excitations in these systems has been shown to depend on the intensity
of the perturbation \cite{calvayrac95}. Here we focus on a different type of external perturbation, namely, that derived from
a point Coulomb charge crossing the system. We will show, nevertheless, that similar shifts in the position of the plasmon peaks are
found. 

Hartree atomic units ($e=\hbar=m_{e}=1$) will be used throughout this work unless otherwise stated.

\section{Methodology}

\
\

Let us first define the system under study: We will focus on a negative point charge (an antiproton) crossing a metal
cluster. 
Electron dynamics in metallic systems typically lie in the femtosecond and subfemtosecond time scales \cite{DMuino2011, Borisov_cpl2004}.
For this reason and for the kind of processes that we study, we consider the cluster ion cores as frozen. We further simplify the problem
and use the spherical jellium model (JM) to represent the cluster.
In the JM, the ions are substituted by an homogeneous background
of positive charge with density $n_0^{+}({\bf r})=n_{0}(r_{s})\Theta(R_{cl}-r)$.
Here $ R_{cl} $ is the radius of the cluster, $ \Theta(x) $ is the Heaviside
step-function and $ n_{0}(r_{s}) $ is the constant bulk
density, which depends only on the Wigner-Seitz radius $ r_{s} $
($1/n_{0}=4 \pi r_{s}^3/3$) ~\cite{Ashcroft}.
The number of electrons in a neutral cluster is $
N =\left( R_{cl}/r_{s} \right)^3$.

The ground state electronic density of the cluster $n({\bf r})$ is obtained
using the Kohn-Sham (KS) scheme~\cite{KS} of density functional theory (DFT)~\cite{HK}.
The ground state KS wave functions $\varphi_{i} ({\bf r})$ are expanded in the spherical harmonics basis set~\cite{Ekardt}.

The evolution of the electronic density in the cluster in response to the field of the moving
charge is calculated
using TDDFT~\cite{RG}.
We propagate the ground state wave functions $\varphi_{i} ({\bf r},0)=\varphi_{i} ({\bf r})$
solving time-dependent KS equations:
\begin{equation}\label{TDKSeq}
\mathrm{i} \frac{\partial}{\partial t} \varphi_{i} ({\bf r},t)=
\left \{-\frac{1}{2} \nabla^2 + V_{eff}({\bf r},t)  \right \} \varphi_{i}({\bf r},t).
\end{equation}
The effective potential includes four terms
$V_{eff}({\bf r},t)=V_{ext}({\bf r},t)+V_{H}({\bf r},t)+V_{xc}({\bf r},t) + V_{\bar{p}}({\bf r},t)$,
where $V_{ext} $ is the external potential created by the positive background.
$V_{H} $ is the Hartree potential created by the electronic density. $ V_{xc} $
is the exchange-correlation potential, calculated with the standard
adiabatic local density approximation (ALDA) with the Perdew-Zunger parametrization of Ceperley-Alder exchange
and correlation potential~\cite{Perdew}. Finally, 
$\displaystyle V_{\bar{p}}({\bf r},t)=-\frac{Q_{\bar{p}}}{\sqrt{(z_{\bar{p}}(t)-z)^2+\rho^2}} \Theta(t)$ is the potential
created by the antiproton and acting on the
valence electrons of the cluster. We use cylindrical coordinates ($\rho, z$) in the time-dependent
calculations, which are more appropriate since the problem has axial symmetry. The origin of coordinates
is located at the center of the cluster.
The antiproton is represented by a negative point charge ($Q_{\bar{p}}= -1$)
which moves with constant velocity $\upsilon$ along the $z$-axis. At time t=0, the antiproton is located
at a distance from the cluster (50 a.u.) far enough to avoid a significant interaction between the projectile and the target. 
The time propagation of the electron wave function is performed
using the time-stepping algorithm:
 $\varphi_{i}(\rho, z, t +dt) = e^{-iH_{i}dt} \varphi_{i}(\rho, z, t)$. The split operator approximation is then used to separate the
potential and kinetic energy terms in the $e^{-iH_{i}dt}$ time propagator.
The action of the kinetic energy operator is calculated using
Crank-Nicolson propagation scheme.
A detailed description of the
numerical procedure can be found in
Refs.~\cite{Borisov1,Borisov-WPP,Chulkov}.

From the time-dependent KS orbitals we obtain the time-evolving electronic density of the excited cluster
$\displaystyle n(\rho, z ,t)= \sum_{i \in occ} |\varphi_{i}(\rho, z,t)|^2$. The force acting on the moving antiproton along
the $z$-axis
is obtained from the time-dependent electronic density and includes the effect of the positive background:

\begin{equation}\label{force}
F_{z}(t) = 2\pi \int\rho~d\rho~dz
\frac{n(\rho,z,t)-n^{+}_{0}(\rho,z)}{[(z_{\bar{p}}(t)-z)^2+\rho^2]^{3/2}}
[z_{\bar{p}}(t)-z].
\end{equation}
The energy loss is then obtained from the integral:
\begin{equation}\label{eloss}
E_{loss}= -\upsilon\int^{\infty}_{0} F_z(t)~dt.
\end{equation}

We will study the energy loss in two different motion cycles. In the first cycle, the antiproton moves towards the cluster with a constant
velocity $\upsilon$, crosses it following a symmetry axis through the cluster center, and eventually moves away until it reaches a turning
point arbitrarily defined. The turning point is at a distance from the cluster far enough not to have any residual interaction. The
cluster is then left in an excited state. The electronic energy transferred to the cluster during the collision is calculated. In the
second cycle, the antiproton 
turns back from the turning point and starts to approach the excited cluster with the same constant velocity $\upsilon$. In the second
crossing of the cluster, the latter now in an excited state, energy is again transferred to the cluster. We calculate the energy lost in
this second cycle and compare the 
obtained value with that of the first cycle.

 \section{Results and discussion}

\
We have chosen a small Al ($r_{s}=$ 2.07) cluster with $N=$18 electrons and with radius $R_{cl}=$ 5.43 a.u. ($\approx 0.29$ nm).
In all the calculations shown in this article, the projectile velocity is $\upsilon=$ 0.5 a.u.
The ALDA-TDDFT method used here predicts very well the energy loss of antiprotons
in Al targets. The method gives very good agreement with measurements in Al bulk for antiproton
velocities up to 1.8 a.u. Above this velocity the excitations of the inner shells in
Al start to contribute to the energy loss and results deviate from the experimental ones~\cite{Quijada1}.

The antiproton starts its motion at time $t=$ 0 from the position $z_{0}= -50$ a.u.
After the first collision, the projectile continues to move until $t=$ 1000 a.u. At this time the second cycle starts and the antiproton
takes the way back to collide again with the cluster. We call $\tau$ to the time interval between both collisions. In both cycles we
calculate the force $F_{z}(t)$ experienced by
the projectile due to the interaction with the cluster through  Eq.~\ref{force}. From $F_{z}(t)$ we obtain
the value of the energy lost by the antiproton $E_{loss}$ by means of Eq.~\ref{eloss}. In addition, we consider three other different time
spots for the second cycle to start: 1003.5, 1005, and 1010 a.u.
The purpose of using different time delays is to check the sensitivity of the final result to the dynamics followed by the electron
density in the cluster excited state.   
With our choice of time delays, the antiproton 
reaches the excited cluster respectively $\Delta \tau=$ 7, 10 and 20 a.u. of time later than in the reference calculation.
Depending on the value of $\Delta \tau$, the antiproton will start to cross the surface of the excited cluster
meeting a minimum or a maximum
in the electronic density oscillations, or an intermediate state. The density oscillations will be discussed later.
The results for the energy loss are summarized in Table~\ref{table}.

The first interesting conclusion that can be extracted from the results of Table~\ref{table} is that the energy loss
of the antiproton crossing the excited cluster is consistently lower
than the corresponding value for the antiproton colliding with the cluster in the ground state.
There are two reasons for the decrease of the energy loss. One reason is that, after the first collision,
the cluster is emitting an amount of electronic charge that roughly corresponds to one electron.
This means that during the second collision
the antiproton is interacting with a smaller amount of electronic charge and therefore loses less energy.
In order to check the relevance of the change in the electronic charge of the cluster, 
we performed an additional calculation, namely that of the energy loss in a positively charged cluster which contains 17 electrons and remains in its ground state.
The obtained value of 0.8211 a.u. is lower than the value of the energy loss in the neutral cluster, which is given
in the Table~\ref{table} and is equal to 0.8527 a.u. The difference between these two results is around 4$\%$.
However, the difference is not as big as in the case of time delays $\Delta \tau=$ 0 and 20 a.u. given in the table,
which is up to 11$\%$ of the value of the energy loss in the first collision.
This allows us to conclude that the emission of one electron from the
cluster only partially explains the observed decrease of energy loss in the second collision.

Another reason for the lowering of the energy loss is that the cluster is excited after the first collision
with the antiproton. Namely, the electronic density of the cluster is starting to oscillate in time with a given frequency.
As we mentioned before, depending on the value of the time delay $\Delta \tau$, in the second collision the antiproton
meets different states of the electronic density oscillations at the surface of the cluster. In what follows we are going
to analyze the difference in the energy loss between two collisions depending on the time delay of the second collision.

\begin{figure}[h!]
 \centering
  \includegraphics[width=0.8\textwidth]{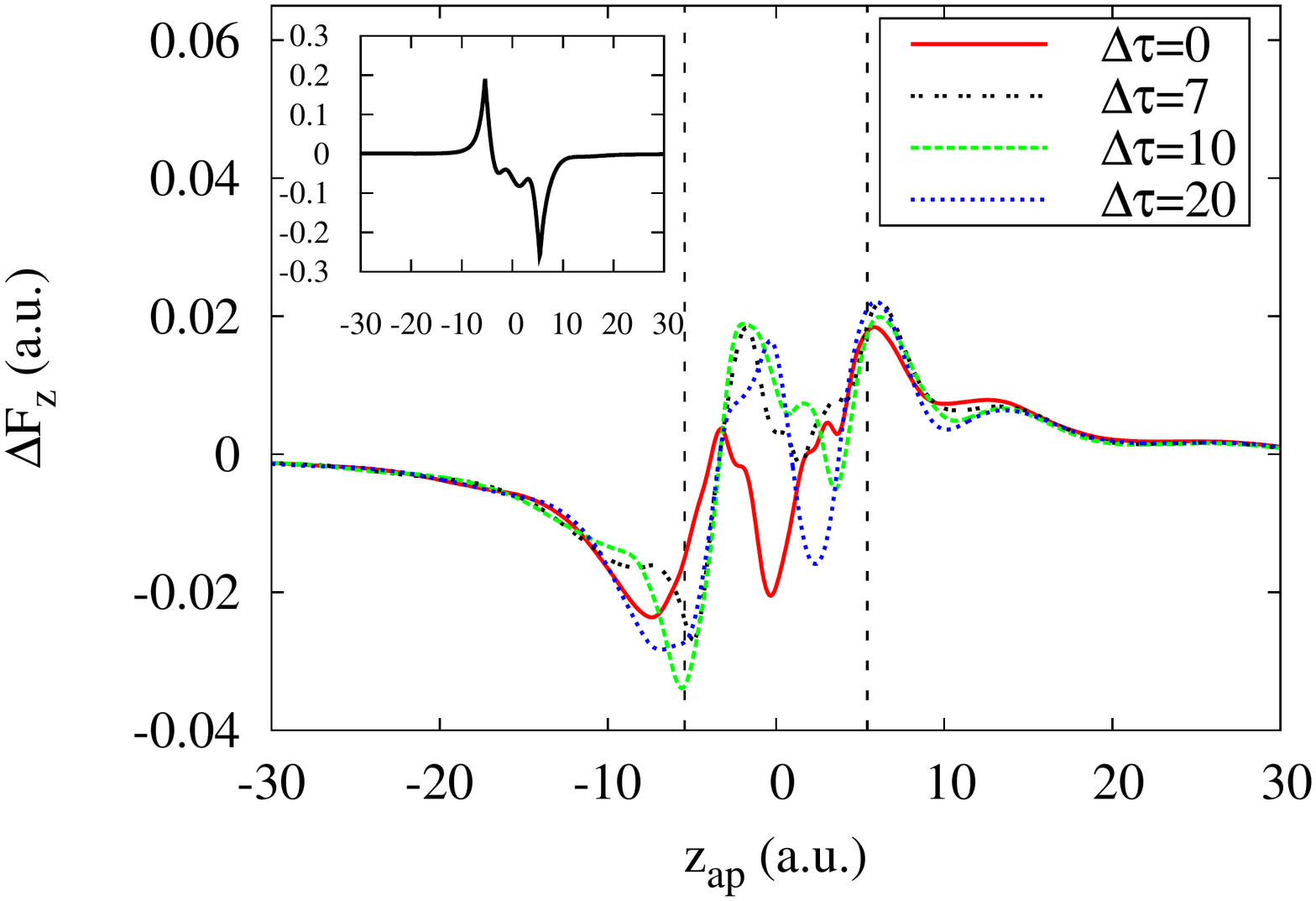}
 \caption{%Difference in the force between the first and the second collision of the antiproton with the cluster, $\Delta F_{z}$,
%for different time delays $\tau$ between the first and second collision. Inset: total force during the first collision $F_{z}^{1st}$, as a function 
%of the projectile position. Dashed lines correspond
% to the radius of the cluster $R_{cl}=$ 5.43 a.u. 
}
\label{dF}
\end{figure}

For different time delays $\Delta \tau$ of the second collision, the value of the energy loss slightly varies.
In order to illustrate this, we show the difference in the force between the first and the second
collision $\Delta F_{z}=F_{z}^{1st}-F_{z}^{2nd}$ for different values of $\Delta \tau$.
Here $F_{z}^{1st}$ is the force felt by the moving charge colliding with the non-excited
metal cluster, which is equal for all $\tau$; $F_{z}^{2nd}$ is the force felt by the antiproton colliding with the excited
cluster. $\Delta F_{z}$ is shown in Fig.~\ref{dF} as a function of the antiproton position.
In this figure, large negative values of $z_{ap}$
indicate positions of the antiproton before each collision with the cluster.
The total force during the first collision $F_{z}^{1st}$ is shown in the inset of Fig.~\ref{dF} as a function of the projectile position.
The two strong features in $F_{z}^{1st}$ correspond to the antiproton crossing the cluster surface. Away from the cluster, the antiproton
is attracted by the induced dipole. Inside the cluster, the electronic density rearranges in order to screen the strong perturbation
created by the moving antiproton. The force inside the cluster oscillates about a mean value that roughly corresponds to the effective
stopping power for this particular velocity of the projectile ($\upsilon=$ 0.5 a.u.)~\cite{Quijada2}.
The curves in the main panel of Fig.~\ref{dF} show how the force felt by the antiproton changes depending on the time at which the second
collision starts.

\begin{figure}[h!]
 \centering
  \includegraphics[width=0.8\textwidth]{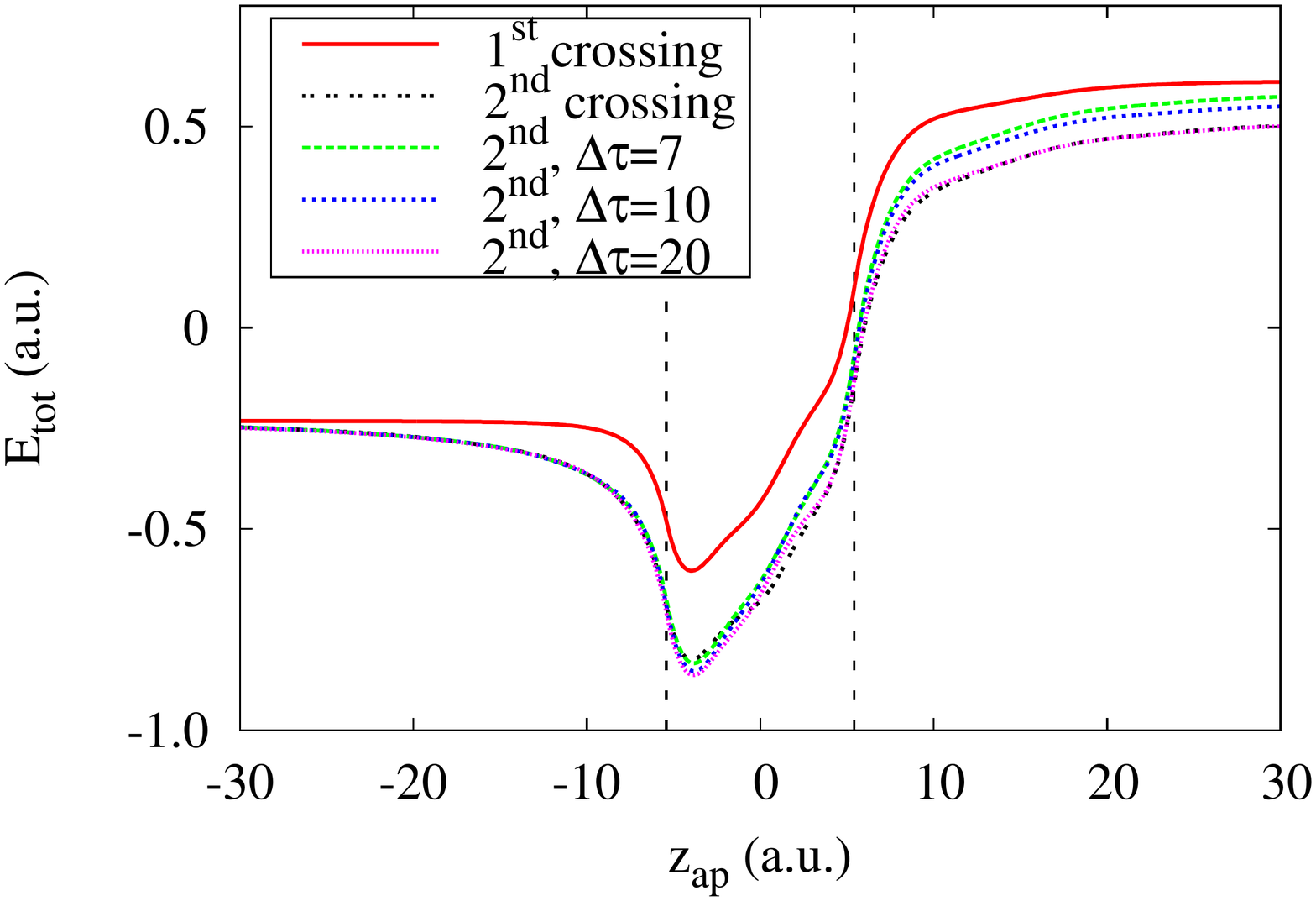}
 \caption{%Total energy of the cluster for different collisions, as a function of the antiproton position.
%All curves corresponding to the second crossing are shifted down on a value of %the energy loss
%corresponding to the first crossing. 
% Dashed lines show the radius of the cluster $R_{cl}=$ 5.43 a.u.
}
\label{Etot}
\end{figure}

The fact that the energy loss is different at different time delays of the second collision can be also analyzed looking at the
total energy of the cluster. The energy is shown in Fig.~\ref{Etot} as a function of the antiproton position.
From Fig.~\ref{Etot} we can see that the total energy of the cluster is increased by the collision. This increase
in energy is the value
of the energy transferred by the antiproton to the cluster or, in other words, the energy lost by the antiproton.
We can see as well that, in all cases, the energy loss after the second collision is lower than after the first collision. 
The curves for $\Delta \tau=$0 and for $\Delta \tau=$20 a.u. are similar. This is consistent with the values
of the energy loss given in Table~\ref{table} for these two cases. We can also see the longer range of the cluster-antiproton
interaction during the second collision. This is due to the net positive charge of the excited cluster.

\begin{figure}[h!]
  \centering
     \subfloat[]{\label{dens-t-cont}\includegraphics[width=0.49\textwidth]{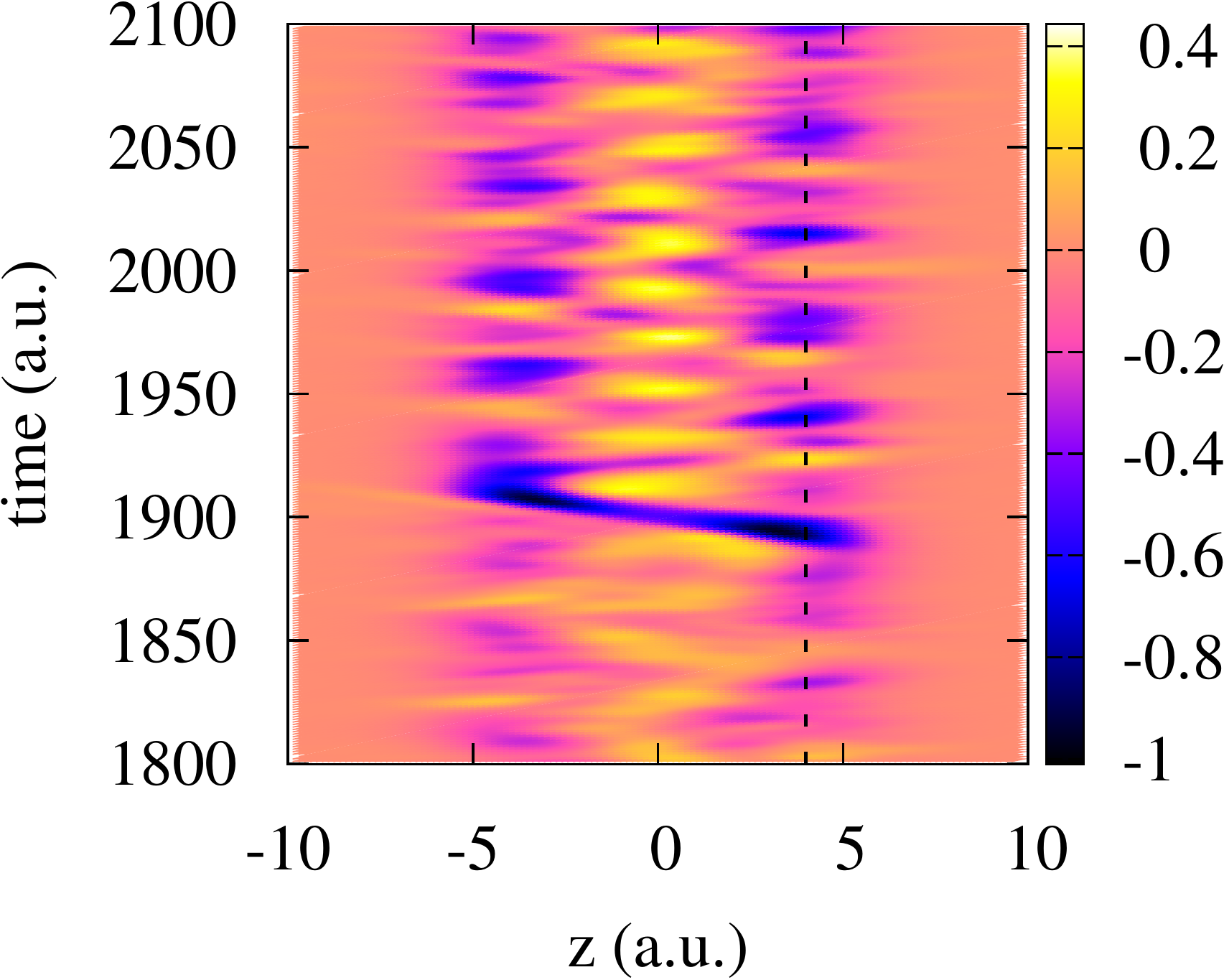}}
     \
     \
     \hfill
     \subfloat[]{\label{dens-t-z-4}\includegraphics[width=0.4\textwidth]{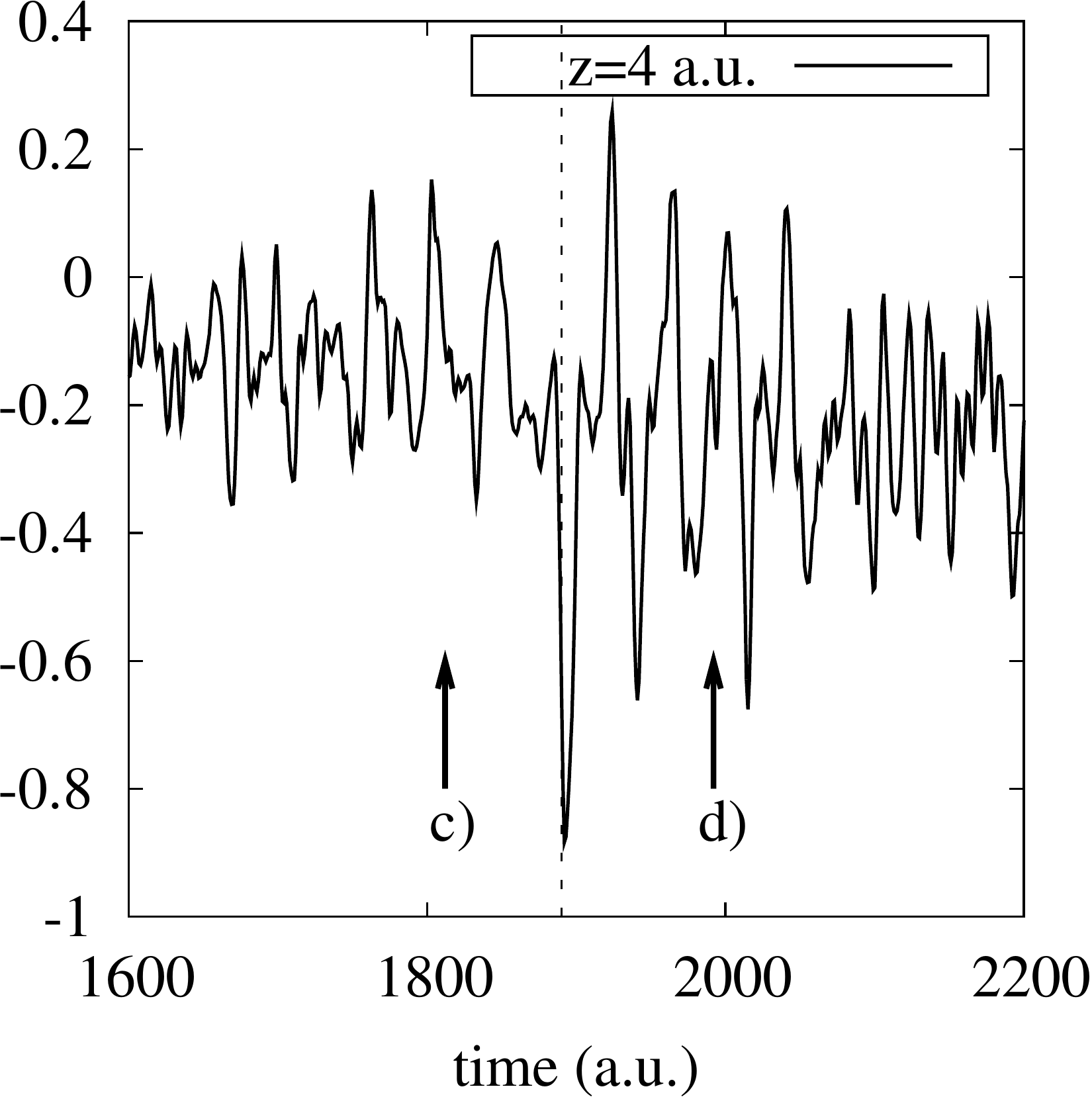}}
     \
     \
     \hspace{17pt}
\hskip 10pt
\\
\hfill
     \subfloat[]{\label{dens-z-ro-1812}\includegraphics[width=0.49\textwidth]{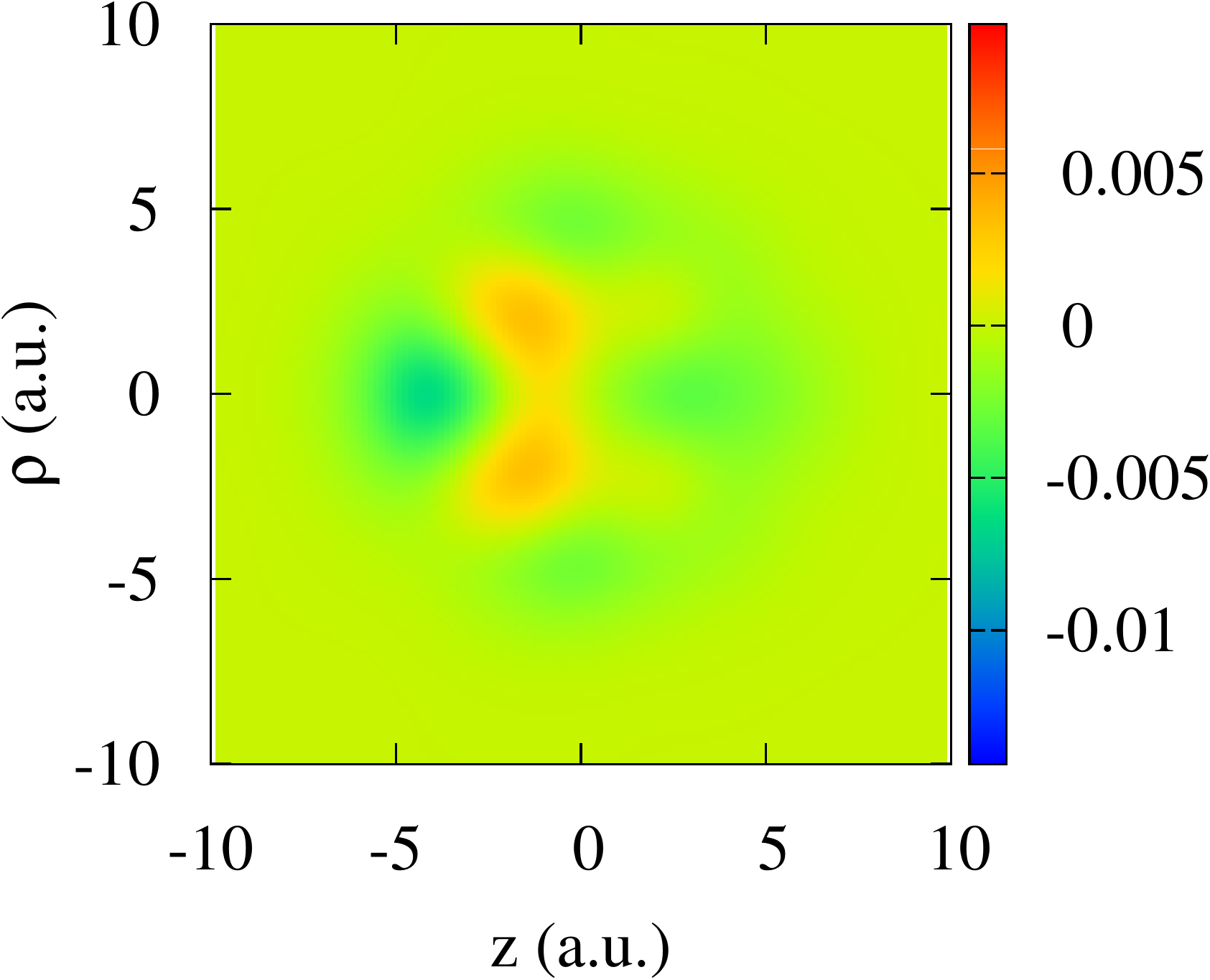}}
     \
     \
     \hfill
     \subfloat[]{\label{dens-z-ro-1992}\includegraphics[width=0.49\textwidth]{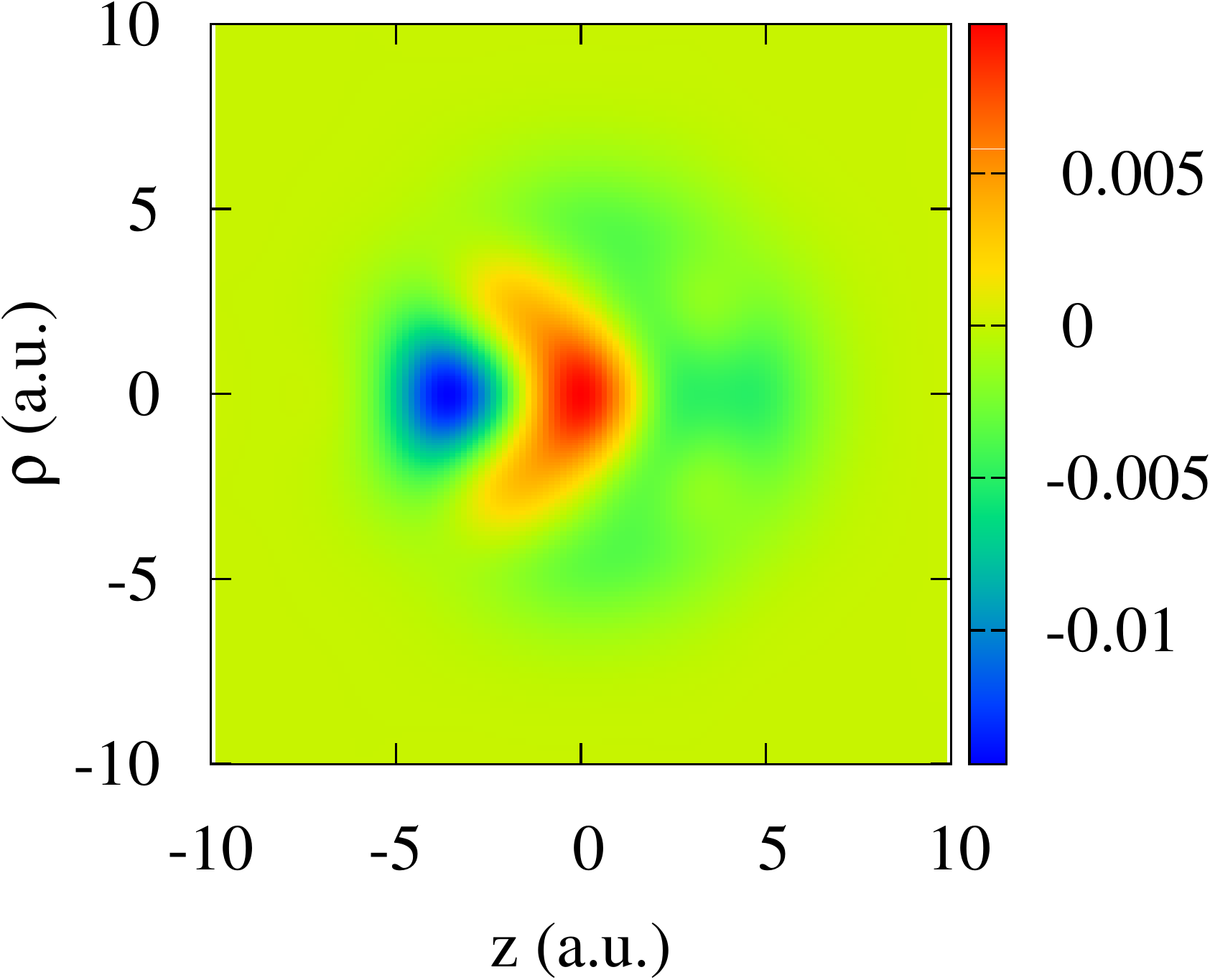}}     
  \caption[]{%(a) Time evolution of the electronic density inside the cluster along the $z-$axis including the time
%  when the antiproton crosses excited cluster. Color code shows the change in density $[n({\bf r},t)-n({\bf r},0)]$
%  in units of the background
%  density $n_{0}$. Dashed line at $z=4$ a.u. indicates the position at which we show (b)
%  the profile of the plot along the time-axis ($\rho=0.02$ a.u.), dashed line corresponds to the moment
 % when the second collision starts. (c) and (d) show the change in the electronic density $[n({\bf r},t)-n({\bf r},0)]$
%  of the spherical cluster (color codes) in ($\rho,z$) coordinates at the time $t=$ 1812 a.u. and $t=$ 1992 a.u. respectively.
}
\label{dens}
\end{figure}

The dependence of the energy loss on the time delay between collisions can be understood by looking at the time evolution of the induced
electronic density. Figure~\ref{dens-t-cont} shows the change in electronic density
$\Delta n(z,\rho=0,t) = [n(z,\rho=0,t)-n(z,\rho=0,t=0)]$ inside the cluster in units of the background density $n_{0}$, along the $z-$axis
and as a function of time.
The results are shown for the calculation with $\Delta \tau=$0. 
The time interval in Fig.~\ref{dens-t-cont} is chosen to include the moment at which the second collision of the antiproton with the
cluster takes place. In the figure, the projectile moves from the right to the left. 
The second collision starts at $t \approx$ 1890 a.u. We clearly see the negative change in density, originated by the Coulomb repulsion
between the incident antiproton and the cluster electrons. From Fig.~\ref{dens-t-cont} we can see that the excitation created by the
moving charge
in the cluster leads to oscillations in the induced electronic density:
Minima and maxima in the induced density are observed. Depending on the time delay between collisions $\tau$, the impact of the incoming
antiproton with the previously excited cluster can bump into a minimum or a
maximum of the electronic density oscillations. 
In the first calculation ($\Delta \tau=$ 0) and when the time delay is $\Delta \tau=$ 20, the antiproton starts
to cross the excited cluster when there is a maximum in the electronic density oscillations at the surface of the cluster
(the change in density in Fig.~\ref{dens-t-cont}
is positive). In the case of $\Delta \tau=$ 10 a.u. 
the second crossing finds a minimum of the electronic density oscillations at the cluster surface.
The time delay $\Delta \tau=$ 7 a.u. is chosen to have a case in which
the second crossing falls neither on the minimum nor on the maximum of the change of the electronic density, but in-between these two situations.
Depending on this circumstance, the value of the energy lost by the antiproton varies.

From Fig.~\ref{dens-t-cont}, we can also see that the minima and maxima in the induced electronic density become more pronounced after
the second collision, indicating that the cluster is further excited by the second collision. 
This can also be seen in Fig.~\ref{dens-t-z-4} where the change in density
is shown as a function of time for the particular value of $z=4$ a.u., marked with a dashed line in Fig.~\ref{dens-t-cont}. 
The amplitude of the electronic density oscillations increases after the second collision. This is also observed
in Fig.~\ref{dens-z-ro-1812} and Fig.~\ref{dens-z-ro-1992} where we illustrate the density distribution in the cluster before
($t=$ 1812 a.u.)
and after ($t=$ 1992 a.u.) the second collision. The change in density is plotted in a plane in ($\rho, z$) coordinates with
the center of the cluster at ($\rho=0, z=$0). The negative and positive peaks in the right panel (Fig.~\ref{dens-z-ro-1992})
are much more intense than in the left panel (Fig.~\ref{dens-z-ro-1812}). These pronounced oscillations show that, after the second
collision, the
oscillations of the induced electronic density are stronger. The excitation created by the
second antiproton enhances that created during the first collision. However,
the similar distribution of the induced charge seems to indicate that
similar electronic modes are excited in both events.

\begin{figure}[h!]
 \centering
  \includegraphics[width=0.8\textwidth]{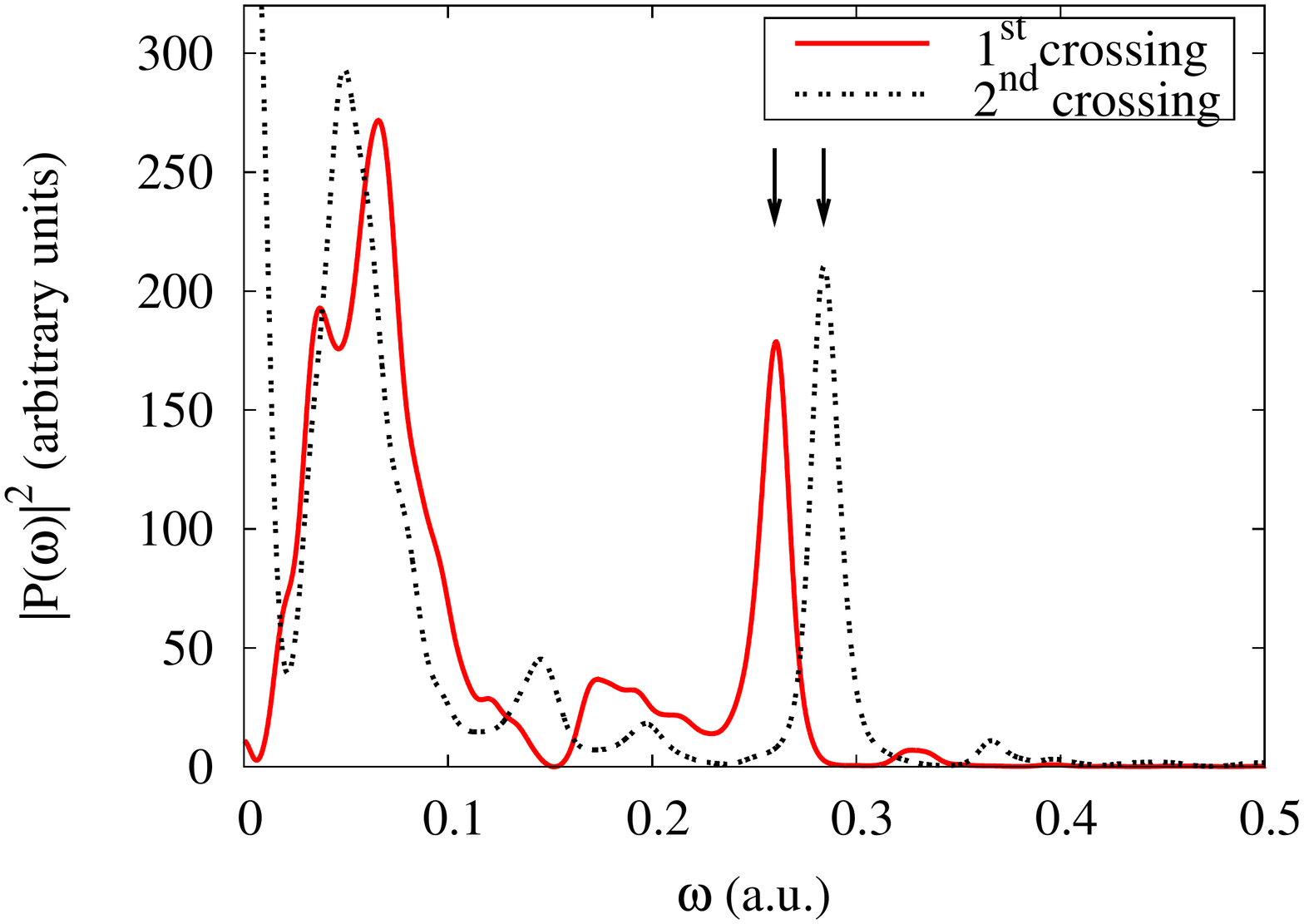}
 \caption{}
\label{dip}
\end{figure}

In order to calculate the frequency of the density oscillations we perform a Fourier analysis of the time evolution
of the dipole moment $P(t) \rightarrow P(\omega)$ created by the electronic density in the excited cluster. The Fourier transform is done
for two different cases: a) after the single collision and without including the second collision, and b) after the two collisions.
In this analysis we use the time evolution of the dipole during $\sim$1200 a.u after
each collision and an exponential mask function (centered in the middle
of such interval) to avoid spurious effects due to the
use of finite time interval.
The results for the respective dipole power spectra 
$|P(\omega)|^2$ are shown in Fig.~\ref{dip}.
Two peaks are shown by arrows at frequencies $\omega=$ 0.261 and 0.284 a.u. (corresponding periods of the plasmon oscillations are
$T\approx$ 24.1 and 22.1 a.u.). They roughly correspond to the expected value of the plasmon energy in the cluster:
The plasmon frequency for a perfect metal sphere can be calculated from the
value of the density $n_{0}$ as $\omega_{p}=\sqrt{4 \pi n_{0}/3}$ (Mie plasmon frequency)~\cite{Blackman}.
For an Al cluster ($r_{s}=2.07$) this value is 
$\omega_{p}=0.34$ a.u. In our calculations the obtained frequency is lower than the frequency given by the
classical Mie theory. This is due to the small size of the cluster
and because we use quantum theory for the calculation of the frequency. A red shift
with respect to its classical Mie value is frequently observed in clusters of simple metals \cite{Kresin,calvayrac2000,deHeer,Brack}.
In Fig.~\ref{dip} we also see that the plasmon peak is shifted to higher frequencies after the second collision.
This behavior is consistent with the non-linear shift of the plasmon frequency under non-perturbative conditions \cite{calvayrac95}.
The blue shift of the frequency after the second collision is in part related to the emission of
electronic charge from the cluster due to the interaction with the antiprotons. It was observed that the resonance position
moves to higher frequencies when increasing the positive charge of the cluster \cite{calvayrac2000}.
Lower-energy excitations are also present in the spectra and can be attributed to excitations of electron-hole pair
character \cite{calvayrac95}.

\
\section{Conclusions}

In summary, we have calculated the energy loss of an antiproton colliding with a small Al cluster, both when the cluster is 
in the ground state and when the cluster is in an excited electronic state. We have shown that the antiproton loses less energy when
penetrating a cluster
previously excited.
The lowering of the energy loss is related not only to the fact that the cluster is transferred to an excited state, but also
to the fact that the cluster loses one electron during the first collision with the antiproton.

We have also shown that the projectile creates a plasmon
in the cluster and that the plasmon peak shifts to higher frequencies in the second collision. This corresponds to the observed shorter
period and larger amplitude of the electron density oscillations in the cluster after the second collision of the antiproton
with the cluster. 
The shift of the plasmon peak to higher frequencies is partially due to the emission of one electron from
the cluster, which thus becomes positively charged.

Our work is another example of how TDDFT in the time domain is an extremely useful tool to study electron dynamics in finite-size objects,
as well as to analyze the energy loss processes of charges interacting with condensed matter.

\section*{Acknowledgements}

NEK acknowledges support from the CSIC JAE-predoc program,
co-financed by the European Science Foundation.
We also acknowledge the support of the Basque Departamento de
Educaci\'on and the UPV/EHU (Grant No. IT-366-07), the
Spanish Ministerio de Econom\'{\i}a y Competitividad
(Grant No. FIS2010-19609-CO2-02) and the
ETORTEK program funded by the Basque
Departamento de Industria and the Diputaci\'on Foral de Gipuzkoa.
\


%% References


\bibliographystyle{model1a-num-names}
\bibliography{<your-bib-database>}






% \end{thebibliography}

\newpage

    \renewcommand*\thetable{\Roman{table}}
\begin{table}[h!]
\begin{center}
\caption{Energy loss $ E_{loss} $ (in a.u.) of an antiproton crossing the spherical Al cluster in ground and excited states.}
  \begin{tabular}{| l|c|c|c|c|c|r }
    \hline
     & $1^{st}$ collision & $2^{nd}$ collision & $2^{nd}$ collision, & $2^{nd}$ collision, & $2^{nd}$ collision, \\ 
     & & & $\Delta \tau=$ 7 au & $\Delta \tau=$ 10 au & $\Delta \tau=$ 20 au  \\ \hline
    $ E_{loss} $ & 0.8527 & 0.7583 & 0.8318 & 0.8099 & 0.7554 \\
    \hline
  \end{tabular}
  \label{table}
\end{center}
\end{table}

\newpage

Figure 1. Difference in the force between the first and second collision of the antiproton with the cluster, $\Delta F_{z}$,
for different time delays $\tau$ between collisions as a function of the projectile position $z_{ap}$.
Inset: total force during the first collision $F_{z}^{1st}$, as a function 
of the projectile position. Dashed lines show the borders of the cluster ($R_{cl}=$ 5.43 a.u.).
All quantities are in a.u. 

\

Figure 2. Total energy of the cluster $E_{tot}$ for different collisions and for different time delays $\Delta\tau$,
as a function of the antiproton position $z_{ap}$.
All the energy curves corresponding to the second crossing are referred to the value of the energy prior to the first crossing when antiproton is far from the cluster.
Dashed lines show the borders of the cluster ($R_{cl}=$ 5.43 a.u.). All quantities are in a.u. 

\
 
Figure 3. (a) Time evolution of the induced electronic density inside the cluster along the $z-$axis ($\rho=0.02$ a.u.)
including the time
  at which the antiproton crosses the excited cluster. The color code shows the change in density $[n(z,\rho=0,t)-n(z,\rho=0,t=0)]$
  in units of the background
  density $n_{0}$. The dashed line in panel (a) indicates the position $z=4$ a.u. for which, in panel (b), we show
  the change in density as a function of time. Dashed line in panel (b) indicates the moment
  when the second collision starts. (c) and (d) show the change in the electronic density $[n({\bf r},t)-n({\bf r},0)]$
  of the spherical cluster (color codes) in ($\rho, z$) coordinates at times $t=$ 1812 a.u. and $t=$ 1992 a.u. respectively.

\

Figure 4. Dipole power spectra $|P(\omega)|^2$ (arbitrary units) for the excited cluster after one collision (red solid line) and after two collisions (black dashed line). Frequency is shown in a.u.

\newpage

\newpage

\newpage
\end{document}